\documentclass[amsmath,amssymb,prl,superscriptaddress]{revtex4}

\usepackage{mathrsfs}
\usepackage{graphicx}
\usepackage{psfrag}

\newcommand{\Lagrangian}{\mathscr{L}}

\newcommand{\beq}{\begin{equation}}
\newcommand{\eeq}{\end{equation}}
\newcommand{\bea}{\begin{eqnarray}}
\newcommand{\eea}{\end{eqnarray}}
 
\def\<{\langle}
\def\>{\rangle}
\def\d{\partial}
\def\+{\dagger}

\newcommand\A{{\mathrm{A}}}
\newcommand\f{{\mathrm{f}}}
\newcommand\B{{\mathrm{B}}}
\renewcommand\L{{\mathrm{L}}}
\newcommand\R{{\mathrm{R}}}

\def\U1A{U(1)$_{\rm A}$}
\def\LambdaQCD{\Lambda_{\rm QCD}}

\begin{document}

\title{Topological Defects in QCD at 
large Baryon Density and/or large $N_c$}
 \author{Ariel~R.~Zhitnitsky}
 
 \address{Department of Physics and Astronomy,
University of British Columbia,
Vancouver, BC, Canada, V6T 1Z1}

%\abstracts
\begin{abstract}
 The main  leitmotiv of the present talk is analysis 
of different topological objects such as strings and domain walls  in QCD. 
As it is well known, the standard model in general (and QCD in particular)
does not support any kind of such objects due to some simple topological arguments.
However, the situation   cardinally changes  when one considers QCD in large $N_c$ limit
or in the limit of large baryon density when the unique light $\eta'$ field
emerges. 
I discuss   $\eta'$- domain walls and $\eta'$ global strings 
which occur in both systems:  in
high density QCD  and in QCD at large $N_c $. 
 I also discuss 
 the effects of quantum anomalies in the presence of topological
 objects  in  high-density QCD.   The
anomaly induced interactions lead to a number of interesting phenomena
(such as the induced currents along the strings)
which may have phenomenological consequences observable in neutron
stars.
\end{abstract}

\maketitle
 \section{ Introduction}
 This talk is based on the recent works \cite{ssz,Forbes:2000et,fzstrings,Son:2004tq}
 devoted to the study of different topological objects in QCD.
 Domain walls and strings are common examples of topological defects
which are present in various field
theories.  Domain walls are
configurations of fields related to the nontrivial mapping $\pi_0(M)$,
while topologically stable strings are due to the mapping
$\pi_{1}(M)$ where $M$ is manifold of degenerate vacuum states.  While
it is generally believed that there are no domain walls or strings in
the Standard Model due to the triviality of the corresponding
mappings, such objects do exist in QCD at large $N_c$
and  in  the regime of high baryon densities, where the symmetries of 
QCD are broken.
The main reason why these two (apparently very different ) regimes
lead to the similar topological objects is very simple: in both cases
the very light $\eta'$ field emerges. 

This talk is organized as follows.  Section 2    will   explain how some  very interesting topological
structures appear in the regime of high baryon density.   Section 3 
is devoted to the analysis of the similar structures   in QCD at large $N_c$.
Finally, Section 4 is devoted to the quantum anomalies in the presence of topological
 objects  in  high-density QCD.  

\section {High Density QCD}
 The rich phase diagram of QCD at high baryon density has attracted
considerable attention recently, see review paper~\cite{Rajagopal:2000wf} for references.
A typical example is QCD with three light
flavors.  The ground state [the color-flavor-locked (CFL) phase] was
determined to break baryon number, chiral and U(1)$_\A$ symmetries. The CFL phase is
characterized by the condensation of
diquark Cooper pairs.  These pairs are
antisymmetric in spin ($\alpha,\beta$), flavor ($i,j$) and color ($a,b$) 
indices:
\beq
\label{diquarks}
  X^{ai}  = \epsilon^{ijk}\epsilon^{abc}\epsilon^{\alpha\beta}
        (q^{jb}_\alpha q^{kc}_\beta)^* ,~~~~
  Y^{ai} = \epsilon^{ijk}\epsilon^{abc}
  \epsilon^{\dot\alpha\dot\beta}
        (q^{jb}_{\dot\alpha} q^{kc}_{\dot\beta})^*. 
\eeq
The following object,
  $ \Sigma=XY^\+ $
in contrast to $X$ and $Y$, is a gauge-invariant order
parameter.
Furthermore $\Sigma$ carries a
nonzero \U1A charge.  Indeed, under the \U1A
rotations 
$
  q\to e^{i\gamma_5\alpha/2}q \, ,
$
 the $X, Y$ fields 
transform as $X\to e^{-i\alpha}X$, $Y\to e^{i\alpha}Y$, and therefore
$\Sigma\to e^{-2i\alpha}\Sigma$.  
\subsection{Effective lagrangian for the $\eta'$ field.}
The color-superconducting
ground state, in which $\<\Sigma\>\neq0$, breaks the \U1A symmetry.  
The Goldstone mode $\eta'$ of this symmetry breaking is described by
the phase $\varphi_A$ of 
$
  \Sigma=|\Sigma|e^{-i\varphi_A} . 
$
Under the \U1A rotation  $\varphi_A$ transforms as
$
  \varphi_A \to \varphi_A + 2\alpha .
$
At low energies, the dynamics of the Goldstone mode $\varphi_A$ is
described by an effective Lagrangian, which, to leading order in
derivatives, must take the following form,
\begin{equation}
  L = f^2 [(\d_0\varphi_A)^2 - u^2 (\d_i\varphi_A)^2] \, .
  \label{Leffnomass}
\end{equation}
   It is well known that the \U1A symmetry is not a true symmetry of
the quantum theory, even when quarks are massless.  The violation of
the \U1A symmetry is due to nonperturbative effects of
instantons.  Since at large chemical potentials the instanton density
is suppressed (see below), the $\eta'$ boson still exists but acquires 
a finite mass.  In other words, the anomaly adds a potential energy 
term $V_{\rm inst}(\varphi)$ to the Lagrangian
(\ref{Leffnomass}),
\begin{equation}
   L = f^2 [(\d_0\varphi_A)^2 - u^2 (\d_i\varphi_A)^2]  -
      V_{\rm inst}(\varphi_A) \, .
  \label{LeffV}
\end{equation}
The curvature of $V_{\rm inst}$ around $\varphi_A=0$ determines the mass
of the $\eta'$.
 Also,  the standard arguments suggest
 that  the potential $V_{\rm inst}$ must depend on the single combination,
  $\varphi_A-\theta$.
Moreover, at large $\mu$, $V_{\rm inst}$ can be found from instanton
calculations explicitly.  
The infrared problem that plagues these calculations in
vacuum disappears at large $\mu$: %the density of 
large instantons are
suppressed due to
Debye screening. 
%, which increases the action of large instantons.
As a result, most instantons have small size 
$\rho\sim \mu^{-1}$ 
and the dilute instanton gas approximation
becomes reliable. One-instanton contribution, proportional to 
$\cos(\varphi_A-\theta)$, dominates in $V_{\rm inst}$. Therefore,
\begin{equation}
  V_{\rm inst}(\varphi_A) = -a \mu^2\Delta^2\cos(\varphi_A-\theta)\, ,
  \label{Vinst}
\end{equation}
where $\Delta$ is the Bardeen-Cooper-Schrieffer (BCS) gap, and
$a$ is a dimensionless function of $\mu$ which can be explicitly calculated\cite{ssz}.
 Here we only note that $a$ vanishes in the limit
$\mu\to\infty$ very fast,  $a \sim
  (\frac{\LambdaQCD}{\mu} )^{b} , ~b=11/3N_c-2/3N_f\, $.
This is an important fact, since it implies
that the mass of the $\eta'$ boson,
\begin{equation}
   m = \sqrt{a/2} \, \frac{\mu }{f} \Delta ,
     \label{meta}
\end{equation}
becomes much smaller than the gap $\Delta$ at large $\mu$.
In this case the effective theory
(\ref{LeffV}, \ref{Vinst}) is reliable, since meson modes  
  much heavier than $\eta'$ 
decouple from the dynamics we are interested in.
 
The Lagrangian (\ref{LeffV}) with the potential (\ref{Vinst}) is just
the sine-Gordon model, in which there exist domain-wall solutions to
the classical equations of motion.  The profile of the wall parallel
to the $xy$ plane is
$
   \varphi_A = 4 \arctan \exp({mz/u}) ,
$
so the wall interpolates between $\varphi=0(z=-\infty)$ and
$\varphi_A=2\pi(z=\infty)$.  The tension of the domain wall is
\begin{equation}
  \sigma = 8\sqrt{2a}\, uf\mu\Delta  
% {4u\over\pi}\sqrt{a}\mu^2\Delta 
  \, .
  \label{sigma}
\end{equation}
 One should note
that in the limit $a=0$ the  spontaneously broken $U(1)_A$ symmetry 
leads to there existence of the global strings\cite{fzstrings} which however
are always accompanied  by the domain walls for any small but
non -vanishing $a\neq 0$. 

\subsection{Decay of the domain wall.}
It is important to understand the mechanism of the decay of the wall.
It has nothing to do with the decay of $\eta'$ meson quanta, 
which are due to $\eta'$ coupling to photons, ungapped 
quarks, or the gluons of the unbroken SU(2)$_c$ subgroup. 
The domain
wall is already a local minimum of the energy, and the decay
of its excitations means only that the fluctuations 
around this minimum, corresponding to deformations
of the wall, are damped. 

The domain wall is not stable because the same ground state
is on both of its sides: $\varphi_A=0$ and $\varphi_A=2\pi$ are equivalent.
The  instability is due to higher energy meson modes integrated
out and not present in the Lagrangian (\ref{LeffV}). One can
visualize the effect of these modes by considering an effective
potential which depends on the magnitude $|\Sigma|$ as well as
on the phase $\varphi$ of the order parameter $\Sigma$.
This potential has the shape of a Mexican hat,  slightly tilted by an
angle proportional to $a$.   
The domain wall is a configuration
that, as a function of the coordinate perpendicular to the wall,
starts from the global minimum, goes along the
valley, and returns to the starting point. One can continuously
deform this configuration into a trivial constant one by pulling
the looplike trajectory over the top of the hat.
This deformation has to be
done in a finite area of the wall first, thus creating a hole.
If this hole exceeds the critical size, it will expand, destroying
the wall. On the rim of the hole the magnitude of $|\Sigma|$
vanishes. The field configuration around the rim is a vortex: on a closed
path around the rim, $\varphi_A$ changes by $2\pi$.  
%Thus the field configuration near the rim is exactly that of a vortex. 
The decay of
the wall is a quantum tunneling process in which a hole bounded by a
closed circular vortex line is nucleated. The
semiclassical probability of this process is\cite{ssz},
\begin{equation}
  \Gamma \sim  \exp\biggl( -{16\pi\over3}{\nu^3\over u\sigma^2} \biggr)
  \, ,
  \label{hole}
\end{equation}
where $\nu$ is the tension of the vortex line in the limit of massless
boson, $m=0$.   

To find $\Gamma$ we   need to compute the string tension $\nu$,
which  to the logarithmical accuracy can be estimated as 
$
  \nu  
      = 2\pi u^2 f^2 \ln (R\Delta) \, ,
$
where $R$ is
%is the radius of the hole, which provides 
a long-distance cutoff (thickness of the wall $\sim m^{-1}$),
and $\Delta$ is the momentum scale at which the effective Lagrangian
description breaks down.  We are helped by the fact that the vortex
tension is dominated by the region outside the core, so the effective
Lagrangian (\ref{Leffnomass}) is sufficient for computing $\nu$ to the
logarithmic accuracy.
 Therefore, the decay rate can be estimated  to be
\begin{equation}
  \Gamma \sim \exp\biggl(-{\pi^4\over3}{u^3\over a}{f^4\over\mu^2\Delta^2}
%  \ln^3{f\over\sqrt{a}\mu} 
  \ln^3{1\over\sqrt{a}}
  \biggr) \, .
  \label{Gamma}
\end{equation}
Since $f\sim\mu\gg\Delta$, and $a$ decreases with increasing $\mu$, the
decay rate is exponentially suppressed at high $\mu$.

\section{QCD at large $N_c$.}

In this section we want to demonstrate that all important ingredients  of the $\eta'$  string 
and domain wall configurations  discussed in the previous section, remain in place for the case
of QCD at large number of colors.
Indeed, the most profound property of QCD at large $N_c$, which
is there existence of the light pseudo Goldstone
$\eta'$ field, remains the same. This is due to the fact that  new small parameter
$1/N_c \ll 1$ enters  into  the system   replacing 
the small parameter   $\Lambda_{QCD}/\mu \ll 1$ discussed previously.
Indeed,  the  $\eta'$   mass in the chiral  limit
$m^2\sim 1/N_c$  vanishes\cite{Witten:1980sp}. This relation replaces the $\eta'$ mass 
formula (\ref{meta}) of the previous section.  Similar to  the previous case, the periodic
properties of the $\theta$ parameter in QCD
 along with the specific dependence of the potential
on the unique combination $(\theta -\varphi_A)$, imply there existence of the $\eta'$ domain wall. 
As in the previous case, the domain wall configuration corresponds to the path which
interpolates between $\varphi_A=0$ and  $\varphi_A=2\pi$ describing the same
physical vacuum states with identical energies (this corresponds to 
the spontaneous breaking of the discrete symmetry,
$\varphi_A=0 \rightarrow \varphi_A=2\pi )$ .
The width of the wall determined by small  $\eta'$ mass. Similar to the previous case,
the spontaneous (rather than explicit) 
violation of the continuous  \U1A symmetry gives rise to there existence of the
$\eta'$ string which is attached to the $\eta'$ domain wall. 

However, there is a difference between the present system and the system discussed in
the previous section. In the case
 of dense quark matter    all calculations were under complete theoretical control for sufficiently 
 large $\mu$.
In the present case of QCD at large $N_c$ while a general $N_c$ counting represents a very robust 
result, some specific properties of the   potential which are responsible for $\eta'$ mass are model dependent. For a specific model for the $\eta'$ potential motivated by the $\theta $ dependence 
in the  form  $\sim\cos(\theta/N_c)$  the $\eta'$ domain wall solution can be explicitly constructed and the corresponding wall tension can be explicitly calculated\cite{Forbes:2000et},
\begin{equation}
\label{sigma}
\sigma=\frac{2 N_c^2}{N_f} f_{\pi}^2m_{\eta'} 
 \left( 1 - \cos \frac{\pi}{
2 N_c} \right)   \sim \sqrt{N_c}
\end{equation}
Similar to the previous section, the domain walls decay due to the 
  the hole nucleation. 
To describe the corresponding process   we should
look for a corresponding ``instanton" which is a solution of Euclidean
(imaginary time, $t=i\tau$) field equations, approaching the
unperturbed wall solution at $\tau\rightarrow \pm\infty$.  In this
case the probability $\Gamma$ of creating a hole with radius $R$  
 can be estimated as follows  
$
 \Gamma \sim e^{-S_0}
$
where $S_0$ is the classical ``instanton" action.
 If the radius of the nucleating hole is much greater than the wall
thickness, we can use the thin-string and thin-wall approximation.
(The critical radius $R_c$ will be estimated later and this
approximation justified).  In this case, the action for the string and
for the wall are proportional to the corresponding worldsheet areas
 \begin{equation}
\label{d4}
S_0=4\pi R^2\nu -\frac{4}{3}\pi R^3\sigma\,.
\end{equation}
The first term is the energy cost of forming a string: $\nu$ is the
string tension and $4\pi R^2$ is its worldsheet area.  The second term
is energy gain by the hole over the domain wall: $\sigma$ is the wall
tension and $4/3\pi R^3$ is its worldsheet volume.  The world sheet of
a static wall lying in the $x$-$y$ plane is the three-dimensional
hyperplane $z=0$.  In the instanton solution, this hyperplane has a
``hole'' which is bounded by the closed worldsheet of the string.
Minimizing~(\ref{d4}) with respect to $R$ we find the critical radius
and the action,
\begin{equation}
\label{d5}
R_c=\frac{2\nu}{\sigma}\,,\qquad
S_0=\frac{16\pi\nu^3}{3\sigma^2}\,.
\end{equation} 
The obtained expression for $S_0$  of course coincides with the one
used in the previous section
(\ref{hole}).
 To complete the calculations we need to estimate the string tension,
 which (as in the previous case) is mainly determined by the region outside the
 string core. The effective largangian description is justified in this region, and therefore,  
 to logarithmic accuracy one gets,
 $
\nu \sim \frac{\pi}{4} N_f
f_{\eta'}^2 \log R  \sim N_c\log N_c,
$
where we have set  
 the outer radius of the string $R$ is the
same order as the thickness of the domain wall $R\sim m^{-1}\sim \sqrt{N_c}$.   
Therefore, we have the
following $N_c$ dependences
\begin{subequations}
\label{eq:Nc-Results}
\begin{align}
\nu &\sim N_c\log N_c\,,&
\sigma &\sim \sqrt{N_c}\,,\\
R_c &\sim \sqrt{N_c}\log N_c\,,&
m^{-1} &\sim \sqrt{N_c}\,,\\ 
S_0 &\sim N_c^2\log^3 N_c\,,&
\Gamma &\sim e^{-N_c^2\log ^3N_c}\,.
\end{align}
\end{subequations}
To conclude:  the $\eta'$ domain walls are  
large classical configurations
with very long life-time $\tau\sim  \exp({N_c^2\log ^3N_c})$.   
We also
notice   that, although both the critical radius for nucleation
$R_c$ and the wall thickness $m^{-1}$ increase, the ratio $m R_c
\sim \log N_c \gg 1$ and the semiclassical approximation used to
derive the decay rate~(\ref{eq:Nc-Results}) becomes justified in the large $N_c$
limit. 

The structure which is just described  is very generic, and we expect that one should be able to 
recover  this structure 
in the string holographic description  of the $\eta'$ at large $N_c$\cite{Barbon:2004dq}. 
One can also argue that 
the qualitative picture described above remains valid 
even for physically relevant case $N_c=3$. In this  case the  domain walls
will have very large life time\cite{Forbes:2000et}, such that they can be even 
produced and studied  at RHIC(relativistic heavy ion collider)\cite{Shuryak:2001jh}. 

\section{Quantum Anomalies in Dense Matter}
 It is well known that
anomalies have important implications for low-energy physics: the
electromagnetic decay of neutral pions is a textbook example.  In this
section I  describe  the roles of anomalies in QCD at finite density.  I follow
our recent work with Son\cite{Son:2004tq}.
We shall see that topological objects discussed previously play a crucial role
in this study.
 The effects coming from these anomalies are
strikingly unusual; they reveal intricate interactions between the
topological objects such as vortices and domain walls, the
Nambu-Goldstone (NG) bosons and gauge fields.  
    
\subsection{Anomalies}
One method to derive anomalous terms in effective theories is to put
the system in external background gauge fields and require that the
effective theory reproduces the anomalies of the microscopic theory.
We thus consider QCD in the background of two U(1) fields: the
electromagnetic field $A_\mu$ and a fictitious (spurion) $B_{\mu}$
field which couples to the baryon current.  Only at the end of the
calculations we will put $B_\mu=\mu n_\mu$, $n_\mu=(1, \vec{0})$,
corresponding to a finite baryon chemical potential.  (The technique
resembles the one used in Refs.~\cite{Kogut:1999iv,Kogut:2000ek} in a
similar context.)  The Lagrangian describing the coupling of quarks
with $B_\nu$ is
\begin{equation}
\label{1}
  \Lagrangian_\B = \bar \psi \gamma^\nu \left(i\d_\nu
    +{\textstyle \frac13} B_\nu\right) \psi
\end{equation}
(the baryon charge of a quark is taken to be $1/3$).  The Lagrangian
is invariant under U(1)$_\B$ gauge transformations
\beq 
\label{qBU1B}
    \psi \rightarrow e^{i\beta/3} \psi,  ~~~~~~~~
     B_{\mu} \rightarrow B_{\nu}+\partial_{\mu}\beta.
 \eeq
Let us assume that the baryon number symmetry is spontaneously broken,
and the condensate is formed.
The low-energy dynamics then contains a NG boson $\varphi_\B$, which
is the U(1)$_\B$ phase of the condensate.  
The transformation property of $\varphi_\B$ can be
found from Eq.~(\ref{qBU1B}),
\begin{equation}
\label{phiU1B}
  \varphi_\B\to\varphi_\B + 2\beta
\end{equation}
where the factor 2 is the baryon charge of the U(1)$_\B$ breaking
order parameter (assumed to have the baryon charge of a Cooper pair of
two baryons).  We also assume the there exist a neutral NG boson which
comes from breaking of a chiral symmetry.  This can be a $\pi^0$,
$\eta$ or $\eta'$ boson.  To keep our discussion general, we introduce
the current which creates this boson,
\begin{equation}\label{JA}
  j^\A_\mu = \frac12
  \sum_{i=1}^{N_\f} Q^5_i\, \bar\psi_i\gamma_\mu\gamma^5\psi_i
\end{equation}
 This current generates a chiral
transformation,
$
  \psi_i \to\exp\Bigl(\frac i2 \alpha Q^5_i\gamma^5\Bigr)\psi_i\,.
$
The NG boson is characterized by a phase $\varphi_\A$, which
transforms under this transformation as
$
  \varphi_\A \to \varphi_\A + q_\A\alpha\,,
$
where $q_\A$ is a number characterizing the axial charge of the
condensate.  The factor $\frac12$ was introduced in Eqs.~(\ref{JA})
  so that the U(1)$_\A$ charge of the chiral condensate
$\<q_\L\bar q_\R\>$ is one.  In color superconducting phases, the
current will be normalized so that $q_\A=2$.

The effective low-energy description must respect the U(1)$_\B$ gauge
symmetry and be invariant under the gauge transformations~(\ref{qBU1B})
and (\ref{phiU1B}).  Therefore, in the effective Lagrangian the
derivative $\d_\mu\varphi_\B$ should always appear in conjunction with
$B_\mu$ to make a covariant derivative $D_\mu\varphi_\B =
\d_\mu\varphi_\B-2B_\mu$.  Vice versa, each occurrence of $B_\mu$ has
to accompanied by a $-\frac12\d_\mu\varphi_\B$.   
 
 In the
presence of the background electromagnetic and U(1)$_\B$ fields, the
conservation of (\ref{JA}) is violated by triangle anomalies:
\begin{equation}
\label{QCD-anom}
  \d^\mu j^\A_\mu = -\frac1{16\pi^2}\bigl(
  e^2 C_{\A\gamma\gamma} F^{\mu\nu}\tilde F_{\mu\nu}
  - 2 e C_{\A\B\gamma} B^{\mu\nu}\tilde F_{\mu\nu}\\
  + C_{\A\B\B} B^{\mu\nu}\tilde B_{\mu\nu}\bigr)
\end{equation}
where $F_{\mu\nu}=\d_\mu A_\nu-\d_\nu A_\mu$ and $B_{\mu\nu}=\d_\mu
B_\nu-\d_\nu B_\mu$; $\tilde F_{\mu\nu} =
\frac12\epsilon_{\mu\nu\alpha\beta}F^{\alpha\beta}$, $\tilde
B_{\mu\nu} = \frac12\epsilon_{\mu\nu\alpha\beta}B^{\alpha\beta}$.  The
coefficients $C$'s in Eq.~(\ref{QCD-anom}) are given by
\begin{equation}\label{Cs}
    C_{\A\gamma\gamma} = 3\sum\nolimits_i Q^5_i (Q_i)^2, ~~~
  C_{\A\B\gamma} =  \sum\nolimits_i Q^5_i Q_i,~~~
     C_{\A\B\B} = \frac13\sum\nolimits_i Q^5_i .
\end{equation}
The effective theory has to reproduce the anomaly
relation~(\ref{QCD-anom}).  In the effective theory $\d_\mu j_\mu^\A$
can be found from the change of the action under the chiral 
transformations: $\delta
S=-\int\!d^4x\,\d^\mu\alpha\, j_\mu^\A =\int\!d^4x\,\alpha\,\d^\mu
j_\mu^\A$.  From the condition of anomaly matching one can deduce that
the effective Lagrangian contains the following terms\cite{Son:2004tq},
\bea
\label{Lanon}
   \Lagrangian_{\rm anom} = 
  \frac1{8\pi^2q_\A} \d_\mu\varphi_\A
  \Bigl[e^2 C_{\A\gamma\gamma} A_\nu \tilde F^{\mu\nu}
    - 2 e C_{\A\B\gamma}
    \left(\mu n_\nu-{\textstyle\frac12}\d_\nu\varphi_\B\right)
    \tilde F^{\mu\nu}    
 - {\textstyle\frac12}
  C_{\A\B\B}\, \epsilon^{\mu\nu\alpha\beta}
  \left(\mu n_\nu-{\textstyle\frac12}\d_\nu\varphi_\B\right)
  \d_\alpha\d_\beta\varphi_\B \Bigr]
\eea 
The term proportional to $C_{\A\gamma\gamma}$ in Eq.~(\ref{Lanon})
describes two-photon decays like $\pi^0\to2\gamma$ and
$\eta'\to2\gamma$.  Such processes occur already in the vacuum.   
We do not consider these terms
here.  The new terms are the ones proportional to $C_{\A\B\gamma}$ and
$C_{\A\B\B}$.
At first sight, these new terms either vanish identically or are full
derivatives and cannot have any physical effect.  This is true when
the NG fields $\varphi_\A$ and $\varphi_\B$ are small fluctuations
from zero.  In particular, $\pi^0$ does not decay to a phonon and a
photon.  However, as $\varphi_{\A,\B}$ are periodic variables, the
action can be nonzero if either or both variables make a full $2\pi$
rotation.  This occurs in the presence of topological defects like
vortices or domain walls discussed in section {\bf 2.}

\subsection { Anomalies in CFL phase}
Our discussion so far has been rather general.  We now specialize
ourselves on   specific color superconducting phase,
 CFL phase, and discuss the physical consequences in the presence of
topological defects in this phase.
In the CFL phase, the gauge-invariant order parameters are 
constructed from diquarks, see eq.(\ref{diquarks}),
 as $\Sigma= X^\+ Y$ which is a $3\times3$ matrix which breaks chiral
and U(1)$_\A$ symmetries.   Obviously, the CFL phase has baryon vortices, see
Ref.~\cite{fzstrings} for details.  In fact, a rotating CFL core
of a star has to be threaded by a vortex lattice.  There are also
U(1)$_\A$ domain walls\cite{ssz} and \U1A strings attached to them(as discussed in Section 2).   
In what follows we consider only few interesting phenomena which by no means
exhaust all possible effects(related to the anomalous lagrangian (\ref{Lanon})).
Our choice simply reflects the author's preferences
and the page constraint requirements for the present Proceedings.

{\em \underline{ Axial current on a superfluid vortex}}.---%
To start, let us consider a superfluid (baryon) vortex.     Let the vortex be parallel to the $z$
axis and located at $x=y=0$.  Around the vortex $\varphi_\B$ changes
by $2\pi$, and in the center of the vortex it is ill-defined.  Outside
the core $\epsilon^{\mu\nu\alpha\beta}\d_\alpha\d_\beta\varphi_\B=0$,
and the term in $\Lagrangian_{\rm anon}$ that contains $\varphi_\B$
vanishes.  However, at the vortex core, as usual,
$(\d_x\d_y-\d_y\d_x)\varphi_\B=2\pi\delta^2(x_\perp)$.  The action
becomes
\begin{equation}
\label{action}
  S_{\rm anom} = \frac\mu{12\pi} \int\! dt\, dz\, \d_z\varphi_\A
\end{equation}
where the integral is a linear integral along the vortex line.  Now
let us recall that the axial current $J^\A_\mu$ is obtained, from
Noether's theorem, by differentiating the action with respect to
$\d_\mu\varphi_A$.  One sees immediately that   there is an axial
current running on the superfluid vortex, with  magnitude 
\beq
\label{J}
J^A=\frac{\mu}{12\pi}.
\eeq   
This  implies that the system induces
a strong classical weak (rather than electromagnetic) field along the vortex.
 
 {\em \underline{Magnetization of axial domain walls}}.---% 
Let us consider an axial domain wall in an external magnetic field.
Such domain walls exist at very high densities where instanton effects
are suppressed as discussed in section 2.  When the baryon field $B_\nu$ is
treated as a background, $B_\nu=(\mu,\vec 0)$, the following term is
present in the anomaly Lagrangian:
\begin{equation}\label{Bgrad}
  \Lagrangian_{\A\B\gamma} = \frac{eC_{\A\B\gamma}\mu}
  {8\pi^2}
  \vec{B}\cdot \vec{\nabla}\varphi_\A
\end{equation}
where we have set $q_\A=2$ and $\vec B$ is the magnetic field.
Consider now a U(1)$_\A$ domain wall streched along the $xy$
directions.  On the wall $\varphi_\A$ has a jump by $2\pi$.  Now turn
on a magnetic field perpendicular to the wall, i.e., along the $z$
direction.  Equation~(\ref{Bgrad}) implies that the energy is changed
by a quantity proportional to $BS$, where $S$ is the area of the
domain wall.  This means that   the domain wall is magnetized,
with a finite magnetic moment per unit area equal to
$eC_{\A\B\gamma}\mu/(4\pi)$.  The magnetic moment is directed
perpendicularly to the domain wall.  For   CFL phase the
magnetic moment per unit area is $e\mu/(6\pi)$.

{\em \underline{Currents on axial vortices}}.---%
The same effect can be looked at from a different perspective.  One
rewrites Eq.~(\ref{Bgrad}) into the following form,
\begin{equation}
  \Lagrangian_{\rm anom} = \frac{eC_{\A\B\gamma}\mu}{4\pi^2}\,
  \epsilon_{ijk} A_i\, \d_j\d_k \varphi_\A
\end{equation}
Since $\epsilon_{ijk}\d_j\d_k \varphi_\A\sim 2\pi\delta^2(x_{\perp})$
on the vortex core, the action can be written as a line integral along
the vortex,
\begin{equation}
   S_{\rm anom} = \frac{eC_{\A\B\gamma}\mu}{2\pi} 
   \int\!d\vec\ell\cdot \vec A
\end{equation}
which means that   there is an electric current running along the
core of the axial vortex, which is similar to the axial current on
baryon vortices.  The magnitude of the current is given by
\begin{equation}
\label{j}
  j^{\rm em} = \frac{eC_{\A\B\gamma}\mu}{2\pi}.
\end{equation}
 Naturally, the electromagnetic current running along a closed vortex
loop generates a magnetic moment equal to $\frac12 jS$, where $S$ is
the area of the surface enclosed by the loop.  This is exactly what we
found in Eq.~(\ref{Bgrad}).  A large vortex loop, therefore, has a
magnetic moment that can be interpreted as created by the current
running along the loop, {\em or} as the total magnetization of the
domain wall stretched on the loop.  The two pictures come from the
same term in the anomaly Lagrangian. The currents flowing on vortices are
reminiscent of Witten's superconducting strings~\cite{Witten}. 

{\em\underline{Axial Current in the Magnetic Field Background $\vec{B}$.}}---%
The effective lagrangian  (\ref{Bgrad})  leads to another
interesting phenomenon\footnote{ I am thankful to Max Metlitski who suggested 
this interpretation of eq.(\ref{Bgrad}).}. Namely, in the quark dense matter in the presence of the magnetic field
$\vec{B}$ the effective lagrangian (\ref{Bgrad})  implies that there is the axial current running
along $\vec{B}$ with the corresponding current density proportional to the magnetic field, 
$\vec{j}^A=\frac{eC_{\A\B\gamma}\mu} {8\pi^2} \vec{B}$. If the system were a type II superconductor,
one could calculate the total current $J^A=\int j^AdS$ carried by a fundamental magnetic flux $\int\vec{B}\vec{dS}=\frac{1}{2e}2\pi$.
 The obtained  current 
 \beq
 \label{J_B}
 J^A=\frac{C_{\A\B\gamma}\mu} {8\pi}  
 \eeq
 is analogous  to the
one calculated previously on the superfluid vortex (\ref{J}). However, the currents
on the magnetic fluxes  (\ref{J_B}) could be much more important phenomenologically 
than the currents on the superfluid vortices 
(\ref{J}) because the density of the magnetic fluxes in neutron stars could be  many orders of magnitude
larger than density of superfluid vortices. 

{\em \underline{Angular momentum  on the axial vortices}}.---%
An axial vortex loop also carries angular momentum.  To see that, let
us place a vortex loop inside a rotating quark matter.  We thus take
 $B_\nu=\mu v_\nu$ where $v_\nu$ is the local
velocity and $\d_i v_j-\d_j v_i=2\epsilon_{ijk}\omega_k$.  The ABB
term in the anomalous Lagrangian leads to the following expression,
\begin{equation}
  \Lagrangian_{\A\B\B} = \frac{C_{\A\B\B}\mu^2}{8\pi^2}
  \vec\omega\cdot\vec\nabla\varphi_\A
\end{equation}
which implies that the domain wall is characterized by a constant
angular momentum per unit area, equal to $C_{\A\B\B}\mu^2/(4\pi)$.  As
in the case of the magnetic moment, this angular momentum can
alternatively be interpreted as being carried by an energy (mass)
current running along the vortex surrounding the wall.
 and the value of the energy
current is
\begin{equation}
  j^{\rm energy} = \frac{C_{\A\B\B}\mu^2}{2\pi}
  = \frac\mu3 \sum_f j_f.
\end{equation}
The last representation is particularly illuminating, since it shows
that the same current (\ref{j}) that creates the magnetic moment  also creates
angular momentum.  We can visualize this current as carried by quarks
with different flavors $f$, where each quark has electric charge $Q_f$
and energy equal to the quark chemical potential $\mu/3$.
  \subsection{Future Directions}
We have seen that the quantum anomalies, especially those involving
the baryon or axial current, lead to new and extremely unusual effects in
high-density matter.   Some effects may have
consequences for the physics of compact objects, which are to be
explored. Of particular interest is the finding that baryon vortices (and magnetic fluxes)
  carry axial neutral current,
which interact with neutrinos and may affect the cooling process of 
neutron stars or/and may influence the neutron star kicks. 
 
All results in this section have been derived from the anomaly terms of
the effective Lagrangian and do not rely on details of the microscopic
QCD Lagrangian.  It would be very interesting  to understand the microscopic origin
of the vortex currents found here.  The first step in this direction  was undertaken quite
recently\cite{Max} where  some  explicit calculations on the microscopical level
have been carried out in a simple model. In particular,
it was shown  that in some cases
  the results derived from the anomaly terms are truly topological and do not get any corrections even
  for finite temperature $T$, while in other cases the corrections  could be of order one.
   
All credits  for this project should go to my close collaborators M.~M.~Forbes, M. Metlitski, E.~V.~Shuryak, D.~T.~Son and  M.~A.~Stephanov. They are not, however,  responsible  for any
 typos, errors and mistakes  unavoidably present in this text.
This work   was supported, in part, by the Natural Sciences and Engineering
Research Council of Canada.

\end{document}